\documentclass[12pt]{article}
\usepackage{amsmath, amsfonts, amsthm, amssymb, color, graphicx, bm, url, xr, zref, comment, setspace}
\RequirePackage[colorlinks, linkcolor=red, citecolor=blue, urlcolor=blue]{hyperref}
\usepackage[top=3.2cm, bottom=3.4cm, left=2.6cm, right=2.6cm]{geometry}

\usepackage[nofiglist, notablist]{endfloat}

\usepackage[square,numbers]{natbib}
 \doublespace

\begin{document}
\title{Bayesian Multi-Arm De-Intensification Designs}

\author{
Steffen Ventz\footnote{\small steffen.ventz.81@gmail.com}, 
Lorenzo Trippa \\
{\small Dana-Farber Cancer Institute  and Harvard T.H. Chan School of Public Health},\\
}

\clearpage
\maketitle
\begin{abstract}
\noindent 
{\small 
 In recent years   new cancer  treatments  improved  survival  in multiple histologies.
 Some  of  these  therapeutics, and  in particular   treatment combinations,  
are often  associated  with severe treatment-related adverse events (AEs).
It  is  therefore  important  to  identify  alternative de-intensified therapies,  
for  example dose-reduced therapies, with reduced AEs  and similar efficacy.
We introduce a sequential design for  multi-arm de-intensification studies. 
The design  evaluates  multiple de-intensified therapies  at  different dose levels,    one  at  the time, based on modeling of toxicity and efficacy  endpoints. 
We  study the utility of the  design in oropharynx cancer  de-intensification studies. 
We use a Bayesian nonparametric model for efficacy and toxicity outcomes to define  decision rules at interim and final analysis. 
Interim decisions  include   early  termination  of  the study due to   inferior  survival  of  experimental  arms compared  to the  standard of care (SOC),  
and  transitions  from  one de-intensified treatment arm to  another  with a  further reduced dose
when  there is  sufficient  evidence  of non-inferior survival.
We evaluate the   operating characteristics of the design using simulations and  data from  recent de-intensification studies in human papillomavirus (HPV)-associated oropharynx cancer.
}

\end{abstract}

%\tableofcontents
%\clearpage

\section{Introduction}
In the last two decades several new cancer treatments  have improved patient survival  \citep{semenza2008new}. 
A large portion of  new therapies consists of a  backbone treatment,  
often chemotherapy or radiation therapy,
combined  with   an additional  drug, for  example a targeted therapy  or  an immune checkpoint inhibitor. 
Some of  these  combination therapies  improved  survival,  
but they are  associated  with  severe  AEs. 
Intensity-modulated radiotherapy (IMRT) in combination with cisplatin 
is a SOC  in  oropharynx cancer \citep{ang2010human} 
with three  year survival rates close to 90\%  \citep{ang2010human, gillison2019}.
However,  the addition of cisplatin to IMRT is associated with a substantial increase in acute and late AEs compared with IMRT alone \citep{munker2001}.
%which adversely impact patients'  lives  \citep{langendijk2008}.  
%
Similarly,   a  combination  treatment including   chemotherapy  is the  SOC for early-stage HER2-positive breast cancer
and  it has  a high rate of  treatment-related AEs \cite{mathew2017less}.
% Several  clinical studies  are currently  evaluating   de-intensified breast cancer therapies with shorter courses of chemotherapy \citep{mathew2016systemic, llombart2017her2}

%{\bf  Definition of  De-intensification studies.} 
AEs  associated with new  anticancer  treatments  are the main
motivation for  testing  if de-intensified therapies  maintain  efficacy similar to  the SOC and  reduce treatment related AEs.
For instance, the  recent studies E1308\citep{marur2017e1308}, OPTIMA\citep{seiwert2018optima}, RTOG1016 \citep{gillison2019}, DeEscalate\citep{mehanna2019}, 
PAMELA \citep{llombart2017her2} and KRISTINE \citep{hurvitz2018neoadjuvant} 
evaluated de-intensified therapies in  oropharynx cancer and breast cancer. 
De-intensification studies consider  therapies that 
(i) are dose reductions of  SOC therapies,
(ii) replace one  component of  the  SOC combination therapy  
with a  potentially less  toxic  drug, or
(iii) eliminate the backbone treatment from the SOC. 
In all  these cases  the clinical study   seeks  to demonstrate that the de-intensified  treatment has  survival  outcomes similar to the SOC and reduces  AEs.

%percentage{\bf Motivation. } 
The design that we introduce is motivated by  a clinical study in HPV-associated oropharynx cancer at  our  institution. 
HPV is a DNA onco-virus  \citep{gillison2000}. 
 HPV positive and negative oropharynx cancer  constitute distinct  cancer sub-types 
with distinct molecular characteristics and epidemiological profiles \citep{gillison2000}. 
% Nontheless,  SOC treatments and dose levels  are nearly identical  in these two populations.
%
%The incidence of HPV-associated oropharynx cancer is increasing  in high-income countries \citep{mehanna2013prevalence}.  
Several ongoing clinical studies are  evaluating  de-intensified therapies \citep{mirghani2018treatment}.
The trial  that  we  designed   will evaluate two de-intensified therapies  which    differ   in the  IMRT dose levels. %(i.e.  radiation intensity). 

Two large de-intensification studies 
RTOG1016\citep{gillison2019} and De-ESCALaTE \citep{mehanna2019}, 
which replaced  {\it cisplatin} with  the EGFR inhibitor {\it cetuximab}, recently reported 
inferior survival   under 
the de-intensified therapy 
 compared to  the SOC (estimated overall survival hazard ratios of 1.45 and 5 for    RTOG1016  and De-ESCALaTE)  without    
 reducing    AEs. 
De-intensified studies in HPV-associated oropharynx cancer tend to use large  margins \citep{Agostino2003} 
for testing non-inferiority 
to reduce   sample sizes at targeted type I/II error rates. 
These  margins and inadequate interim analyses   
can lead - as the  results of RTOG1016 and De-ESCALaTE suggest - 
to a  large  number of patients  exposed  to  treatments  with reduced  efficacy. 
%
%The design of 
This indicates  the  importance   of 
sequential  de-intensification designs to    handle trade-offs between   
power 
%(null hypothesis:  the de-intensified treatment has inferior survival compared the the SOC or does not reduce AEs), 
and  the   %control  of  the  
number  of    patients    exposed  to  inferior and toxic treatments using adequate  interim analyses (IAs).

%{\bf Bayesian de-escalation designs.} 
We introduce  a  Bayesian design for   de-intensification studies.  
The   design  allows investigators  to   test    multiple  treatments sequentially, 
for instance  two de-intensified treatments with $80\%$ and $40\%$ of the original IMRT dose of the SOC. 
Using a  Bayesian nonparametric model for the distribution of survival times %, either overall  survival (OS) or progression free survival (PFS),
and AEs
we specify     sequential  decision rules   to evaluate treatment response and toxicity reductions.
%   in  de-intensified therapies.
These  include early stopping rules
  that  are tuned to balance
(i) the  risk of   patients receiving inferior treatments that  reduce   survival and 
(ii) the need to identify non-inferior  de-intensified  treatments that  reduce  toxicities. %  with  sufficient  power.
The multi-arm design evaluates  de-intensified  therapies one at the time starting  from dose-levels  close to the SOC, 
with  subsequent arms at lower dose-levels tested  only  if there  is  evidence  of  non-inferiority  and reductions of AEs
for    the previous de-intensified treatments.  

We discuss  algorithms to tune stopping rules accordingly  to pre-defined early stopping probabilities
under the null hypothesis of inferior survival or  AEs identical  to  the  SOC.  
Additionally the  design    calibrates the type I error rate to approximately match  a targeted $\alpha$-level. % at a pre-defined set of null-scenarios.
We evaluate the operating characteristics of the design  
using %simulation scenarios  selected using 
data from   recent de-intensification trials in HPV-associated oropharynx cancer.
%\footnote{when do you tell reader  if there is  a control or not?  early in sec2 is fine}

%{\bf Prior work:}
De-intensification designs use
non-inferiority (NI) testing procedures   with a pre-defined NI margin  
and evaluate if the efficacy of an experimental treatment is comparable to the SOC  \citep{Agostino2003}.  
Statistical considerations for NI  studies concern 
the selection of a suitable testing procedure, 
the specification of the  NI margin $\Delta$, 
and the study design,  including early stopping rules and the selection of the sample size  \citep{blackwelder1982, rothmann2003, freidlin2007, joshua2012, Korn2017a}.
\cite{blackwelder1982}   discussed NI tests  based on asymptotic techniques and
 \citep{farrington1990, tu1998, laster2006}  focused on the finite-sample operating characteristics  of   NI tests.
Exact NI tests  have been discussed in  \citep{chan2003, laster2006},
and   extensions to time-to-event outcomes 
have been proposed in \citep{rothmann2003, freidlin2007}.
Other contributions   focused on the  selection of suitable NI margins $\Delta$ \citep{snapinn2004, holmgren1999}
and on the specification of  early stopping rules for sequential NI experiments \citep{freidlin2002comment, lachin2009futility, Korn2017a}.

 Bayesian work  on NI experiments  includes   NI testing methodologies and 
 the use of   data from previous  clinical studies  in the analysis of  NI experiments  \citep{Simon1999, S2013}.
 \cite{W2005} and \cite{ PeppleWilliamson2007} discussed NI tests for binary endpoints using beta prior distributions,  
and \cite{Osman2011} proposed the use of Bernstein  priors.  
\cite{Gamalo2011} used  Bayesian modeling  to select the NI margin $\Delta$
and \citep{Daimon2008, Chen2011} investigated Bayesian sample size calculations for  single-stage NI tests. 

The main difference  between these non-sequential single-stage non-inferiority testing procedures  
and our work is that we introduce a sequential design for de-intensification studies with efficacy and toxicity  co-primary endpoints.
 The    design utilizes  a non-parametric Bayesian model  to  analyze   survival data and AEs during the study.
%At interim analyses, 
Key decision to pause, stop or  continue the evaluation of  de-intensified treatments, 
are  based  on  data summaries that  quantify  the  trade-off  between the risk  of  exposing patients  to an  inferior  treatment and  the likelihood of demonstrating relevant reductions of AEs.

The outline of the paper is as follows. 
After introducing some notation in Section \ref{Sec:Design},
we present the de-intensification  design for  studies with efficacy endpoints (Section  \ref{Sec:Design:EfficacyEndpoint}) and for studies with efficacy and toxicity co-primary endpoints (Section \ref{Sec:Design:Two:Endpoints}).
Section \ref{Sec:Prior} summarizes the Bayesian probability model that we used. 
In Section \ref{Sec:Simulation:Study:Effiacy} we evaluate the operating characteristics of the trial design under different design  parameters.
Section \ref{Sec:Simulation:Study:Effiacy:Parametric}   compares several  de-intensification  strategies 
in HPV-associated oropharynx cancer with efficacy endpoints. 
Section \ref{Sec:Simulations:Effiacy:Tox} extends this  comparison 
to oropharynx cancer studies with efficacy and toxicity co-primary endpoints.

\section{De-intensification design}\label{Sec:Design}
%\subsection{Preliminaries}\label{Sec:Notation}

%
We consider   a  phase II clinical study with   $k=1, \cdots, K$  de-intensified treatments. 
We  assume  the  study  does  not  include  a  control arm.   
%and discuss necessary
 Simple modifications of the   design  that we  discuss    allow  to  include a control arm. % in section Section \ref{Sec:Disuccion}. 
%%\footnote{SV: Add Section}.  
In the de-escalation setting, the  SOC  ($k=0$)  survival distribution  has  been  estimated previously
and the SOC is  associated  with a substantial risk of AEs.  %\footnote{SV: No meaning. All  cancer treatments  are associated  with  AEs }
De-intensified treatments $k=1, \cdots, K$ are  likely  to  present  better toxicity profiles 
(i.e. reduced doses of the backbone treatment), 
but may  reduce patients' survival. 
In practice $K=2$ or $3$. $K=2$ in our study at DFCI. % and our simulations below $K=2$. 

We evaluate if one or multiple treatments $k$ are  non-inferior compared to the SOC and reduce AEs.  
De-intensified treatments 
 are  ranked.
For example,  if $k= 1, \cdots, K$  are de-intensified dose levels,
 the study  starts testing   the highest  dose ($k=1$), followed   by further dose reductions  $k=2,\ldots,K$.

%\subsection{Evaluation Treatment response}
A maximum of $n$ patients are enrolled.
For each patient  $1 \leq i \leq n,$  $(T_i, C_i , Y_i, X_i) $  indicates the assignment  to arm $C_i \in  \{1, \cdots, K\}$ at enrollment time $T_i\geq 0$,
%\footnote{no $T_i$ in (...) there is $E_i$} 
and   $Y_i$ and $X_i$ are  the  efficacy and toxicity outcomes. 
In our  study, $Y_i$ indicates the progression free survival (PFS) time and  $X_i$ is the time of  the first 
treatment-related AE (grade $\geq$3). 
%\footnote{SV: AE adverse events, defined above}
 $F_k(\cdot)$ and $G_k(\cdot)$ indicate  the distributions  of  $Y_i$ and $X_i$ for patients on  treatment $k$,
and $n_{t,k}$ and $n_t$  are the number of enrollments to arm $k$ and total  number of enrollments at time $t$. 
Lastly, $\Sigma_t$ 
%= \{ (Y_{i}(t), R_{Y,i}(t), X_{i}(t),  R_{X,i}(t), C_i ) \} _{  i :  T_i < t   }$, %\footnote{should be $C_i$  not $C_\ell $ }
%\footnote{$\Delta$: notation conflict.  $\Delta$  has a  different meaning in this  paper. Use $R$  for  censoring. }
%\footnote{$C_il$????? $C_i$  second  time  that  i correct this}
denotes  the  data collected until time $t$ since the first enrollment.%,
%where 
%$
%Z_{i}(t)=\min(Z_i, t-T_i)  
%X_{\ell}(t)=\min(X_\ell, t-T_\ell)
%$  
%and 
%$R_{Z,i}(t) = I(Z_i \leq t-T_i)$ are  censored versions of $Z=Y, X$. 
%
%Lastly  $n_{t,k}$ and $n_t$  are the number of enrollments to arm $k$ and total  enrollments by time $t$. 

In some cases  AEs reductions  of  treatments $k$ compared to the SOC  
can be anticipated or have been demonstrated  
before the de-escalation study, 
while in other cases it is necessary to estimate both toxicity and efficacy during the trial. 
Therefore we first  introduce  in Section \ref{Sec:Design:EfficacyEndpoint} a design for clinical studies that utilize  only survival  outcomes for interim  and  final  decisions,    
and then  extend in Section \ref{Sec:Design:Two:Endpoints} the design  to include  both toxicity and efficacy co-primary endpoints. 
 The designs can be combined  with any  Bayesian model 
 %$p( F, G)$ \footnote{same  notation F ... different meanings - not good -  futility  or distribution.  here cut  $p( F, G)$.
%i%n the MC part  if you use $F_i\in \mathcal{F}_I$  as  you suggested  yesterday  you  create  a  different confusion with  $F_k$. maybe  $\tilde F_i$ or  $\tilde F$
 %}
 for the unknown  distributions $F_k(\cdot)$ and $G_k(\cdot)$.  

%\footnote{The  use  of $\Delta$  in expression 1  and  6  is  inconsistent: different meaning and  pourpose. use $\Delta_H $  in 6  and  change  table1}

\subsection{De-intensification studies with efficacy primary outcomes}\label{Sec:Design:EfficacyEndpoint}

For each de-intensified treatment $k$, efficacy   is quantified by a summary  $\theta_k = \theta(F_k) \in \mathbb R$.
A large $ \theta_k$  corresponds  to a large treatment effect.
Examples  include  
the median,  or 
the restricted mean survival time (RMST)  $E[\min(Y_i, t_E) |  C_i=k, F_k ]$ at a pre-specified $t_E>0$. 

For each  $k=1, \cdots, K$ the null and alternative hypotheses that we consider are
\begin{align}\label{Hypotheses}
\mathcal H_{0,k}=  \{\theta_k  \in \mathbb R: \theta_k \leq \theta_0 - \Delta \}  ~~ \text{ and } ~~
\mathcal H_{A,k}=  \{\theta_k  \in \mathbb R:  \theta_k > \theta_0 - \Delta_{k}\}.
\end{align}
Here $ \Delta \geq \Delta_{k} > 0$ are pre-specified  margins.  
Values  of $\theta_k$ below $\theta_0 - \Delta$  make arm $k$  inferior,
whereas $\theta_k \ge \theta_0 - \Delta_{k}$ indicates an attractive alternative to the SOC.
%The  parameter   $\Delta_{k}$  has an impact  on power   and can be  chosen based  on  prior information about  toxicity reductions,
%with $\Delta_{k}$  proportional  to  the  expected reduction of severe AEs.
The  design  evaluates treatments  sequentially (we will consider $K=2$), one after  another, 
starting with arm $k=1$,  which   is less   likely  to  be inferior than the others.
% Indeed the $k=1,2$  are ranked dose levels of the same  treatment combination.
%(highest dose in  dose de-escalation studies).

\begin{figure}[hp]
	\center 
\includegraphics[scale=.96065] {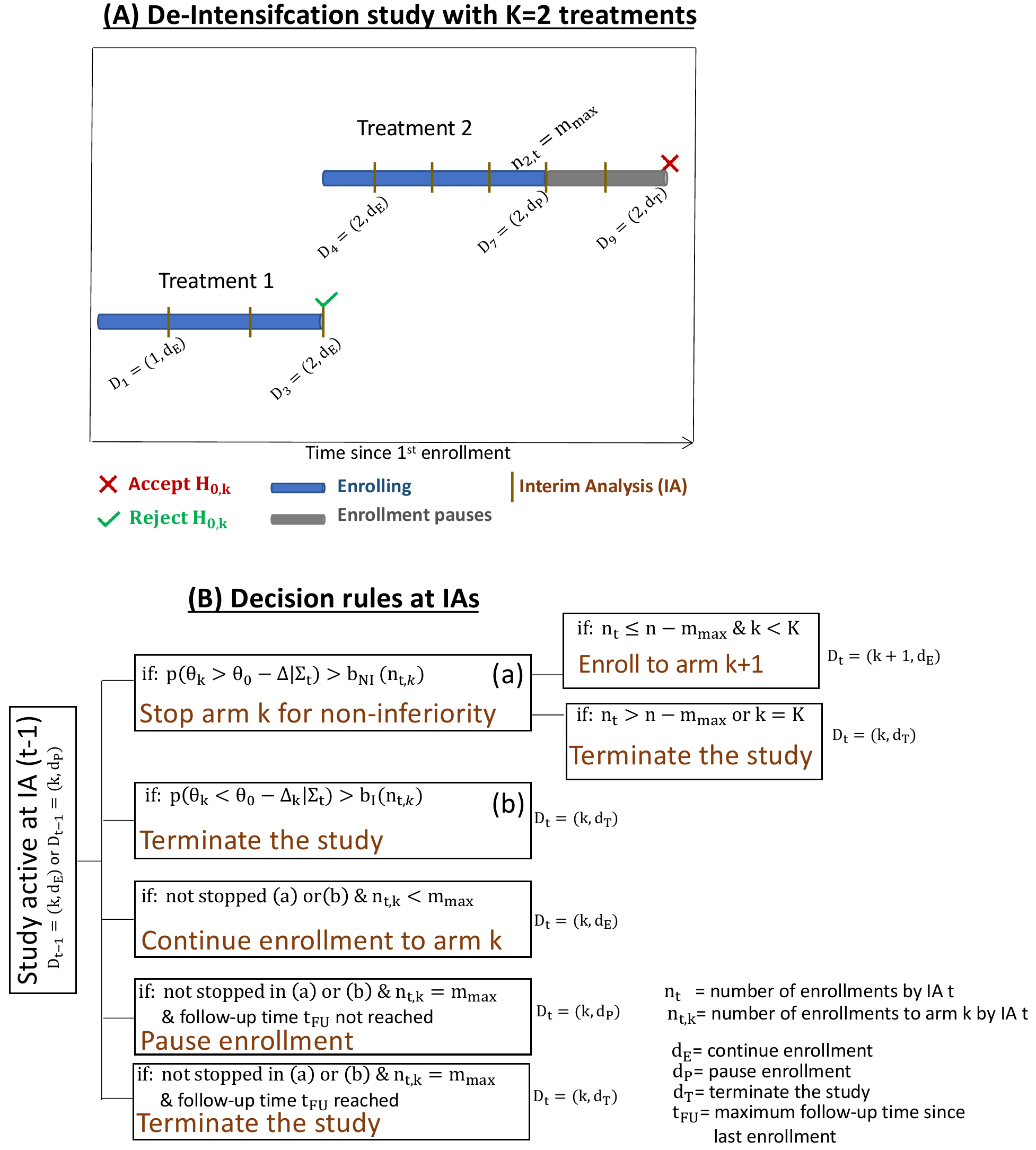}	
\caption{Schematic representation of the trial design with  $K=2$ treatments.
See Section \ref{Sec:Design:Two:Endpoints} for the integrated use of efficacy and toxicity co-primary endpoints.  }\label{Fig:Design1:Scheme}
\end{figure}

Figure \ref{Fig:Design1:Scheme}  
illustrates the design  and a list of possible  decisions  at  interim analyses (IAs). 
A maximum of $m_{\max} \leq n$ 
%\footnote{SV: I will leave $m_{index}$ for all max and minimum sample size requirements, else danger of introducing inconsistent notations via changes}
 patients will be  assigned to each  treatment and 
a minimum of $m_{NI} \leq  m_{\max}$ 
enrollments to treatment $k$ are required before it  can be declared non-inferior.
At regular time intervals   $t=1, 2, \cdots, $ (e.g.  monthly  or quarterly)
IAs 
are conducted, and the active  experimental arm $k$   may be (Panel B of Figure \ref{Fig:Design1:Scheme})
\begin{itemize}
\item[(i)] declared non-inferior to the SOC, and  the  trial progresses  to evaluate treatment $k+1$,
\item[(ii)] declared inferior to the SOC and the study terminates, or
\item[(iii)] enrollment to treatment $k$ continues, or  
\item[(iv)] enrollment  is paused for a maximum    follow-up time $t_{FU}$ since the last enrollment to arm $k$, to allow the accumulation of sufficient information for  testing  $\mathcal H_{0,k}$. 
\end{itemize}
%
%
%\footnote{figure size  fixed; big people can read; Dont make it smaller}

{\small \begin{table}
\begin{tabular}{|l|l|}
\hline
\hline
Parameter                                  &   \\
\hline
\hline
%$F_k, \theta_k$                     & Efficacy distribution and summary for treatment $k=1,\cdots, K$ \ \\
%$G_k, \beta_k$                      & Toxicity distribution and summary for treatment $k=1,\cdots, K$\\
%$n_{t,k}, n_t$                         & number of enrollments to arm $k$ and  overall enrollments by IA $t$ \\
$m_{\max}$                           & maximum number of enrollments per arm \\
%& \\
\hline
$m_{I}, m_{T}, m_{NI}$        & minimum  number of enrollments  to treatment k  before  arm $k$ can be stopped \\
                                              &  for inferiority ($m_{I}$) or toxicity ($m_{T}$), and $\mathcal H_{0,k}$ can be rejected  ($m_{NI}$) \\
%     &                                         \\
\hline
$\Delta, \Delta_L, \Delta_k$  & margins used for testing non-inferiority  and for early futility stoping   \\
%  &                                         \\
\hline
$\Delta_\beta$                       &    margin used for early stopping due to insufficient toxicity reductions    \\
%$b_{NI}(\cdot), S_{NI}, s_{NI}, p_{NI}$   & \\
%$b_{F}(\cdot),  S_{F}, s_{F}, p_{F}$       & \\
%  &                                         \\
\hline
$b_{j}(\cdot), S_{j}, s_{j}$     &   non-inferiority ($j=NI$), inferiority $(j=I)$ and toxicity  $(j=T)$ boundary $b_j(\cdot)$, \\
                                             &   with shape and scale  parameters $S_{j} $ and  $s_{j}$  \\
%                                               &                                         \\
\hline                                             
$p_{I}, p_{T}$                     &   $s_{j}, j=I, T$ are selected so that a proportion of $p_I$ and $p_T$ trials are  \\
                                             &   stopped early at IAs when treatment $k$ has inferior survival ($p_I$) \\
                                             &   or does not reduce toxicities $(p_T)$\\
\hline
\hline
\end{tabular}
\caption{Summary of parameters in the de-intensification designs with efficacy primary endpoint  (Sections \ref{Sec:Design:EfficacyEndpoint}),  
and efficacy and toxicity co-primary endpoints  (Sections \ref{Sec:Design:Two:Endpoints}).}
 
\end{table}

}

Let $D_{t} =( D_{t,1}, D_{t,2})$ indicate 
the de-intensified treatment  $D_{t,1} \in  \{\emptyset, 1, \ldots, K \}$  
that is evaluated  between  IAs  $t$ and $t+1$ ($\emptyset$  if the study is terminated at IA $t$), 
and   $D_{t,2} \in   \{ d_E, d_P, d_T\}$  denotes  the status of the study during this time interval, 
where   
$d_E=$"{\it enrollment  is open}",
$d_P=$ "{\it enrollment is paused} ", 
and $d_T=$  "{\it the study has been  terminated"}  (Figure \ref{Fig:Design1:Scheme}).

{\it \underline{Stopping  rules}:} 
Let  $b_I (\cdot)$ be a predefined futility stopping-boundary  with $1 \geq  b_I(n_{t,k})  \geq 0$  (see  Section \ref{calibration1} for  details). 
Let $D_{t-1}=(k,d_E)$ between IA $t-1$ and $t$. 
%where $k=1$ for simplicity, %\footnote{cut "where $k=1$ for simplicity,"  it decreases  simplicity}  
Arm $k$ is stopped for futility at IA $t$
if the  probability of low  efficacy  becomes larger than $b_I(n_{t,k})$,  i.e. if
%If the predicted probability of low efficacy for arm $k$ becomes larger than $b_0(n_{t,k})$,  
%i.e. if
\begin{align}\label{Stop:H0}
 p( \theta_k \le \theta_0 - \Delta_{k} | \Sigma_t)  > b_I(n_{t,k}),
\end{align}
%arm $k$ is stopped for futility,  
and the study is terminated,  $D_t=(\emptyset, d_T).$

If  there  is  evidence  of non-inferiority 
\begin{align}\label{Stop:HA}
p( \theta_k > \theta_0 - \Delta  |  \Sigma_t)  > b_{NI}(n_{t,k}),
\end{align}
arm $k$ % \footnote{why $=1$} 
is declared non-inferior to the SOC. 
Here  $0 \leq b_{NI}(n_{t,k})\leq 1$
 is a  pre-specified non-inferiority stopping boundary.
If there is evidence of non-inferiority according to (\ref{Stop:HA}), %\footnote{wrong equation quoted} 
the study  proceeds to  treatment  $k+1$ 
conditionally on the availability of a  sufficient   sample size $n_t \leq n-m_{\max}$, i.e. $D_t = (k+1, d_E)$.
Otherwise,  the study is terminated,  
$D_t = (\emptyset, d_T)$. 

{\it  \underline{Pause enrollment:} } 
If  the  probabilities in (\ref{Stop:HA}) and  (\ref{Stop:H0})   don't cross the stopping boundaries, then
enrollment to arm $k$  continues, $D_{t} =(k,d_E)$, unless the maximum enrollment per arm $m_{\max}$ has been   reached. 
When $n_{t,k}=m_{\max}$, 
%\footnote{SV: as indicates above, I will leave all $m_{\cdots}$ as is}
 enrollment is paused, i.e.
 $D_{t} =(k , d_P)$,  
until treatment $k$ is  declared non-inferior or inferior according to  (\ref{Stop:H0}) and  (\ref{Stop:HA}) at later IAs, 
or until the follow-up  time  $t_{FU}$  %\footnote{p} 
 since the last enrollment is reached. 
This potential pause   is necessary   before  testing  arm $k+1$
to limit   patients'  exposure to inferior treatments.
      %
%     \footnote{i  dont  know  what  happen  at  the end  of $t_{FU}$  if  no boundary is crossed.  stop trial ?  is it stated?} 
If the  probabilities in (\ref{Stop:HA}) don't cross the stopping boundaries by the end of the follow-up period $t_{FU}$, the study closed and the null hypothesis $\mathcal H_{0,1}$ is not rejected. %de-intensified treatment is use not 
  %
%by the by during the 
%suring by time
%fuuntill the by t      
%If enrollment  was  paused between IA $t-1$ and $t$, $D_{t-1}=(k,d_P)$,
%and the  probabilities in (\ref{Stop:H0}) and  (\ref{Stop:HA}) didn't cross the boundaries, 
%then   enrollment remains paused, i.e  $D_{t}=(k,d_P),$  and   
%after  the  end of the  follow-up time  $t_{FU}$ the study terminates and     $\mathcal H_{0,k}$ is  not  rejected.
%The evaluation of the remaining  arms $k\geq 2$ is identical.

\subsubsection{Calibration of the design  thresholds}\label{calibration1}
We use  functions $b_j(\cdot), j=I$ or $NI$, of the form
 %that decrease from 1 to  $(1-s_j) \in [0,1]$ with $n_{t,k}$ %to arm $k$
 %
\begin{align}\label{Design1:boundary}
b_{j}(\ell) =  1-s_j  \times  \max  \bigg[ 0,    \dfrac{\ell -m_{j}  } {m_{\max} - m_{j}  }  \bigg]  ^{ S_{j} } \text{ for } \ell=1, 2, \cdots, m_{\max}. 
\end{align}
The parameter  $S_j\geq 0$
determines the shape of $b_j(\cdot), j=NI,I$, 
which is   decreasing from 1 to  $(1-s_j) \in [0,1]$
when $S_{j}>0$  and constant $b_j(n)=1-s_j$ for $m_{j} \leq n \leq m_{\max}$ when $S_{j}=0$.
Here  $m_{NI}$  and  $m_I$ 
indicate the minimum number of enrollments 
necessary before  $\mathcal H_{0,k}$ can be  rejected  and   before  early futility stopping.
%
%Non-increasing  boundaries ($S_{j}\geq 0$) are desirable - they  
%require large posterior probabilities for stopping early during the clinical study  
%when only a few efficacy outcomes have been observed.   
%%The parameter $S_{j}$  determines the shape of $b_j(\cdot)$ , which is   decreasing when $S_{j}>0$  and constant  when $S_{j}=0$.%) boundaries.
%\footnote{i suggest  to cut, not smooth english "Non-increasing  boundaries ($S_{j}\geq 0$) are typically desirable, 
%which larger posterior probabilities to declare non-inferiority (inferiority) early during the clinical study 
%when few efficacy events have been observed. "}

We fix   $\mathcal S=(s_I, S_I, S_{NI}, m_{NI}, m_{I})$ 
%\footnote{$s_F, S_F$??? should be $s_I, S_I$}
(see Section \ref{Sec:Simulation:Study:Effiacy} for a discussion on the selection of these parameters), and 
calibrate the  parameter $s_{NI}$ of the  boundary $b_{NI}(\cdot)$ that  bounds the type I error rate
at the desired  $\alpha$ level
across a    set $\mathcal F_{I}$ of inferior (I) scenarios 
 that satisfy  $\theta( F) =  \theta_0-\Delta$    for each $ F \in \mathcal F_{I}$. 
%\footnote{did we introduce the notation $\theta(\cdot)$?}
%
%
%{\it Definition of $\mathcal F_{I}$.}

{\it Scenarios:} We use the historical control $F_0$ (published Kaplan-Meier estimator) and  select a set of transformations $F= g(F_0)$ (proportional hazards, accelerated failure time, proportional odds, etc.) such that $\theta( F) = \theta_0-\Delta$ for each $F$ in $\mathcal F_{I}$. 
%\footnote{Bad  notation :  (i)  I  contains in $\mathcal F_{I}$ multiple points (ii)  conflict  with $F_k$.  change  to  $\tilde F$}

%{\it Monte Carlo estimate of $s_{NI}$.}
{\it Controlling the type I error rate:}  
For each $ F \in \mathcal F_{I}$,
we  determine 
the largest value of  $s_{NI,  F}$  that 
bounds the designs' type I error rate for arm 1 at level $\alpha$ when $F_1=F$.
%\footnote{ $s_{NI, F}$   new  notation, not  defined. probably unnecessary.  replace with $s_{NI}$ }
%  such that  
%the  design with  parameter  $ s_{NI}= s_{NI,  F} $  (the  remaining $\mathcal S$  parameters are fixed) bounds
%for   $k=1$ the type I error rate at level $\alpha$ when $F_1=F$.
%
%\begin{align}\label{Type:I:Eq}
%\mbox{P}_{F_I} \Big(  \mathcal H_{0,k} \text{ is  rejected} 
%\bigcup_{ \substack{t: D_{t-1,1}=1,\\ ~ D_{t-1,2} \neq d_T }} 
 %\!\! \!\! \!\! \!
%\Big \{   p( \theta_1 \geq \theta_0 - \Delta_{I}  | \Sigma_t)  \geq b_{NI}(n_{t,1}); F_1=F   \Big \} 
%\bigg) \leq \alpha.
%}
%\end{align}
%
We then  set 
$ s_{NI} = \min_{ F \in \mathcal F_I}  s_{NI,  F}.$  
%\footnote{SV: Min here and Max s such that TI below alpha is correct.}
Relevant operating characteristics, 
such as power and the average study duration, 
depends on the selected $\mathcal S,$
and in Section \ref{Sec:Simulation:Study:Effiacy} we discuss the selection of these parameters.

{\it Calibration.}  We estimate $s_{NI,  F}$ using a Monte-Carlo procedure, 
by simulating $C$ trials  (we use $C=2000$ in Section \ref{Sec:Simulation}) 
with  individual outcomes  generated from  $ F$  and random  enrollment times
(with a fixed  enrollment rate) of $m_{\max}$ patients. 
For each simulation $c=1, \cdots, C$, 
we compute the number of enrollments $n^{(c)}_{t,1}$ by  IA  $t=1, 2, \ldots,$ 
and the posterior probabilities $U^{(c)}_{I,t}$ and $U^{(c)}_{NI,t}$ to declare inferiority ($U^{(c)}_{I,t}$) in (\ref{Stop:H0}) and non-inferiority ($U^{(c)}_{NI,t}$) in (\ref{Stop:HA}). 
%using the simulated data of patients with  enrollment by  time $t$. % in dataset $c$ .
%\footnote{cut "using the simulated data of patients with  enrollment by  time $t$. "}
%
Since  $U^{(c)}_{NI,t} > b_{NI}(n^{(c)}_{t,1})$  in (\ref{Stop:HA}) is equivalent to $s_{NI} < s^{(c)}_{t}$, where
$$s^{(c)}_{t}= \Big( 1-U^{(c)}_{NI,t} \Big)  \Big /  \max  \bigg[ 0,    \dfrac{ n^{(c)}_{t,1} -m_{NI} } {m_{\max} - m_{NI}  }  \bigg]  ^{ S_{NI} },$$
the simulated trial $c$ does not reject the null hypothesis $\mathcal H_{0,1}$ (at  time $t$, or at  any other interim analysis) if  $s_{NI}$ is larger or equal than 
$\displaystyle s^{(c)}_{NI}  = \min s_t^{(c)}$ where the minimum is over all $t$  such that  $U^{(c)}_{I,t'} < b_{I}( n^{(c)}_{t',1} )  \text{ for  } t'    \leq t$.
%\footnote{CUT ${ t :   U^{(c)}_{F,t'} < b_{F}( n^{(c)}_{t',1} )  \text{ for  } t'    \leq t  }$  a  punch to the reader.
%maybe add "Here  the minimum is over  the   IAs   that  evaluate for  non-inferiority  arm $k=1$."}
%
%\footnote{equation problems :  (i)  should  be a  minimum  not  a maximum. 
%should  be  "$=\min_{t}\Big \{ s_t^{(c)}\}$  where $t$  ranges  over  the  IAs  
%of  the  trial,  until  the  futility  threshold  is reached $(p^{(c)}_{F,t} < b_{F} \big(n^{(c)}_{t,1})$  
% or until completion of  the  follow up  period $t_{FU}$."
% }
We  then estimate $s_{NI,  F}$   as the $\alpha$-percentile of   $\{s^{(c)}_{NI}\}_{c=1}^C$.

%Since the prior (and posterior) distributions are restricted to the subspace  $\mathcal M_\theta,$ 

%If  the   $\mathcal H_{0,1}$, or  more  generally   $\mathcal H_{0,k-1}$, 
%is  rejected   and  the  trial  proceeds  to  the    arm  $k$ the parameter  $s_{NI}$  is re-estimated  before  starting   enrollment   to arm  $k$. 
%Since $\mathcal H_{0,k}$ is  only tested    after the rejection of    $\mathcal H_{0,k-1}$, 
%the overall family-wise type I error is  preserved at  the $\alpha$ level.
%As    $\mathcal H_{0,k+1}$ is only tested when   $\mathcal H_{0,k}$ was rejected (i.e. a gatekeeping rule\cite{westfall2001optimally, bauer1998testing} with test $\mathcal H_{0,k}$ as gatekeeper the $\mathcal H_{0,1+k}$ test),  

%Because    $\mathcal H_{0,k+1}$ is only tested if  $\mathcal H_{0,k}$ was rejected (i.e. a gatekeeping rule\cite{bauer1998testing}),  
%the overall family-wise type I error rate of the study is bounded below  $\alpha$.

\subsection{Efficacy and toxicity co-primary outcomes}\label{Sec:Design:Two:Endpoints}
When little is  known about AEs of the  de-intensified treatments, 
it  becomes  necessary to evaluate   AEs   together with efficacy as  co-primary endpoints.
%In other instances there is a need to formally evaluate, for each therapy, reductions in  toxicities  compared  to the SOC.
%We extend the design of  Section \ref{Sec:Design:Efficacy:Only} and to   toxicity and efficacy co-primary endpoints. 
%
%Different toxicity endpoint can be considered,  
Recall that $X_i$ indicates  the time (months since enrollment)   of the first  AE (grade $\geq 3$) for  patient $i$, 
with distribution function $G_k$ for arm $k$,
 and $\beta_k$ is a toxicity summary.  
Small   values of   $\beta_k$ indicate  high toxicity.
 In  Section \ref{Sec:Simulations:Effiacy:Tox}  we  use   the RMST $\beta_k = E[  \min(X_i, t_E)  | C_i=k, G_k ]$. 

We consider the null and alternative hypotheses  %that we want to test are 
\begin{align*}
\mathcal H_{0,k} &=  \{ (\theta_k, \beta_k)  \in \mathbb R^2 :  \theta_k \leq \theta_0 - \Delta \text{ or }  \beta_k \leq \beta _0 \},  \text{  and  }  %\label{Hypotheses:2:H0} 
\\
\mathcal H_{A,k} &=   \{ (\theta_k, \beta_k)  \in \mathbb R^2 :  \theta_k > \theta_0 - \Delta \text{ and }  \beta_k > \beta _0 + \Delta_\beta \},  %\label{Hypotheses:2:HA}
\end{align*}
where $\Delta_\beta>0$, and extend the design in  Section \ref{Sec:Design:EfficacyEndpoint}  to include toxicity outcomes.

%de-intensified treatments $k\ge2$ are only tested after  $\mathcal H_{0,k-1}$ has been rejected.
{\it \underline{ Stopping rules}:} 
Treatment $k$, at  time $t \geq 1$,   is declared  non-inferior  and less  toxic than the SOC   
if the posterior probability of  $\mathcal H_{A,k} $ crosses a pre-specified  boundary $b_{NI}(n_{t,k})$, i.e. 
$$p\big( \{ \theta_k > \theta_0 - \Delta \} \cap \{ \beta_k > \beta_0  \} |   \Sigma_t \big)  \geq b_{NI}(n_{t,k}).$$
The function $b_{NI}(\cdot)$ and the futility boundaries introduced below belong to  the same  parametric family  (\ref{Design1:boundary}). %has the same and the remaining are provided below.
When   $\mathcal H_{0,k} $   is  rejected  the trial proceeds to enroll patients to  arm $k+1$ 
if the sample size $n_t \leq n- m_{\max}$, 
 % ($n_t > n- m_{\max}$) 
%\footnote{you  told  me  $m_{\max}$  not $m_{\min}$?  double-check?}  
%is below $n- m_{\min},$ %patients have been enrolled into the study, 
%$D_{t}=(k+1,d_E)$, 
otherwise  the study terminates.%, $D_{t}=(\emptyset, d_T)$.

We    extend   the  early  termination  rules  for inferiority to  include toxicities.  
Specifically, arm $k$ is stopped due to  insufficient early  evidence  of  toxicity reductions 
if the posterior probability of the event
$\{ \beta_k \le \beta_0 + \Delta_{\beta}  \}$, for  $\Delta_{\beta} \in \mathbb R $, 
exceeds the  toxicity boundary $b_{T}(n_{t,k})$, 
%\begin{align}\label{Stop2:H0:A}
i.e. 
$p\big( \beta_k \le \beta_0 + \Delta_{\beta}  | \Sigma_t\big)  > b_{T}(n_{t,k}).$
%
%
%Since $\beta_{k+1} \geq \beta_k$ with probability one, it is reasonable to avoid stopping arm $k\geq 2$ for toxicity after $\mathcal H_{0,1}$ was rejected, 
%and we therefore set  $b_{T}(n_{t,k})=1$ for $k>1$.%

%{\bf Stopping  for  inferiority.}  
Additionally, 
therapy $k$  can  be stopped for inferiority, if the posterior probability of the event $\{ \theta_k \le \theta_0 - \Delta_{k} \}$ becomes larger than the  boundary $b_{I}(n_{t,k})$. 
 The margin $\Delta_{k}= \Delta_{k}(\Sigma_t)$   now depends on the current evidence of toxicity reductions 
%for simplicity we omit the dependents $\Sigma_t, t$ and additional parameters in (\ref{ET:Stop2:H0}) below
i.e.  a  function  of the available toxicity  data, and can vary during the study. 
In particular, with low  evidence of toxicity reductions we use a smaller %larger \footnote{should be  "reduced"  not "larger"}
margin than in  
presence  of  strong  evidence of relevant  toxicity reductions,
%, and we require thereby more efficacy-evidence to continue treatment evaluation when toxicity reductions are rather unlikely.  
%
\begin{align} 
\Delta_{k} (\Sigma_t)= 
\begin{cases} 
 \Delta      & \text{ if }       p\big( \beta_k \le \beta_0 + \Delta_{\beta}  | \Sigma_t\big)   \in (0,  B_T(n_{t,k}) ],  \\
 \Delta_L  & \text{ if }       p\big( \beta_k \le \beta_0 + \Delta_{\beta}  | \Sigma_t\big)  \in (B_T(n_{t,k}),   1],  \label{ET:Stop2:H0}
%1 & \text{ if }  p\big( \beta_k \le \beta_0 + \Delta_{\beta}  | \Sigma_t\big) > b_{T}(n_{t,k}),
\end{cases}
\end{align}
where $ \Delta \geq   \Delta  _L,$
%\footnote{SV:  There is no need for an extra $\Delta_H$ other than $\Delta$ here, and limits  the number of extra design parameters }
$B_T(n_{t,k}) < b_T(n_{t,k}),$  and
$B_T(n_{t,k}) $ is a  parametric  function of the form   (\ref{Design1:boundary}). 
%The 1 in the definition of $\Delta_{k} (\Sigma_t)$ 
%is arbitrary and does not impact the design, because the study terminates  when  $ p\big( \beta_k \le \beta_0 + \Delta_{\beta}  | \Sigma_t\big) > b_{T}(n_{t,k})$. 
%\footnote{the  sentence "Here the function  (\ref{ET:Stop2:H0}) can take  value 1, because the study terminates  when  $ p\big( \beta_k \le \beta_0 + \Delta_{\beta}  | \Sigma_t\big) > b_{T}(n_{t,k})$. " has no logic.  and reader  try to  understand that 1  well before  that  sentence $=$  difficult  to follow}
The  arm is stopped early for inferiority if  the posterior probability
%\begin{align}\label{Stop2:H0:B}
$p\big( \theta_k \le \theta_0 - \Delta_{k}   | \Sigma_t\big)$  
exceeds the   boundary $b_{I}(n_{t,k}).$ 
%In summary, using (\ref{ET:Stop2:H0}), we reduce the decrease the inferiority margin    
%when the early  toxicity data are not promising.

{\it  \underline{Pause enrollment:} } 
Similar to Section \ref{Sec:Design:EfficacyEndpoint},
if the posterior probabilities don't  cross the stopping boundaries,
the study continues
enrollment to  arm $k$  until  $m_{\max}$ enrollments to arm $k$ is reached. 
If  $n_{t,k} =  m_{\max}$ 
enrollment  is  paused and  IAs are conducted until the  maximum follow-up time $t_{FU}$ is reached.

{\it \underline{Calibration of  decision rules}:} 
%\subsubsection{Calibration of Stopping Rules}\label{Sec:Calibration2}
%{\bf Calibration of  decision rules.} The calibration of the stopping rules is similar to  Section \ref{calibration1}.
We  first specify  $b_T(\cdot),$ $B_T(\cdot),$ $b_I(\cdot)$ and the shape parameter of $b_{NI}(\cdot)$. We  then  
determine the parameter $s_{NI}$ of $b_{NI}(\cdot)$ that approximately
%) \footnote{very  unnecessary $()$}
bounds the type I error  over a finite set  of null scenarios $( F,  G)$ using an algorithm nearly identical to Section \ref{calibration1}.
%\footnote{as  earlier:  is I an index  in $(\widetilde F_I, \widetilde G_I)$? }
  %
We  include  in the set of null scenarios 
two extreme cases: % \footnote{SV: Add why this cases are sufficent}

 \noindent (i)  a degenerated toxicity distribution $G$ without AEs  in combination with distributions $F$  
% footnote{use  $\tilde F$, conflict of notation}
  such that $\theta(F)=\theta_0 - \Delta$, %\footnote{use  $\tilde F$, conflict of notation}
 where we consider again transformations  $F=g(F_0)$ of the  SOC   Kaplan-Meier estimate  $F_0$, 
 and 
 
\noindent  (ii)  the case $F=F_0$ %\footnote{use  $\tilde F$, conflict of notation}
 (we assume that the de-intensified treatment can not  improve  survival  compared to the SOC) 
and  distributions $G$ such that $\beta(G)=\beta(G_0)$.
%\footnote{here  you need to be  in the  null  boundary.  i  guess  it  should be $\beta(G)=\beta(G_0)$. }
 %equal to the historical controls (Kaplan Meier estimates).
%\footnote{what  is $\widetilde G_0$?  here  we  want  a beta  value  slightely  better (lower value)  than the SOC.  
%Otherwise  overall  the  design can promote  treatments  with negligible AE reduction and  Delta  loss  in survival.
%i think  the AE reduction has  to be relevant, not just better than zero}

Based  on    monotonicity  relations that  link the  outcome distributions $(F,  G)$   
and the sequential decisions,    
if   $s_{NI}$  bounds  the type I error below   $\alpha$  
in  the outlined settings (i) and (ii),  then $s_{NI}$    
bounds the type I error below  $\alpha$   also under   any other  scenario  within $\mathcal H_{0,k}$.

\section{Prior Probability Model}\label{Sec:Prior}
%One may choose a parametric or a non-parametric Bayesian model for $\{ (F_k,G_k) \}_{k=1}^{K}$.
%For  $\{ (F_k,G_k) \}_{k=1}^{K}$ one may choose a parametric or a non-parametric model. 
In our oropharynx cancer study
we considered several parametric models,
but  based  on available  prior data we  observed unsatisfactory model fits and decided to use a non-parametric prior. 
Both  $F_k$ and $G_k$ are  random  survival  functions with independent prior distributions. 
%$p_G$ and $p_F$  \footnote{notation conflict:  cut  $p_G$ and $p_F$  }

We use a Beta-Stacy  (BS) prior  \citep{walker1997}  for $F_k$ and $G_k$. 
The prior  is strictly related  to the  Dirichlet and  Beta  processes  \citep{ferguson1973bayesian, hjort1990nonparametric}.
For  $W \sim  BS( W | V_0, c)$, where $W$ is either $F_k,$ or  $G_k$,
the distribution  $V_0(t)= E[W(t)]$ is  the  prior  mean    %\footnote{misleading notation -  null hypothesis}
 and  the continuous function $c(t) >0$  controls   the  variability  of    $W$.  
 %\footnote{notation conflict, no H around. }
%a particular neutral-to-the-right (NTR) prior process \citep{doksum1974}.  prior \citep{ferguson1973bayesian, doksum1974}  
%
Under the BS prior,   $\{ -\log(1-W (t)) \}_{t\geq 0}$   is  a monotone, right-continuous random function with 
independent increments,   
 $W (0)=0$ and  $\lim_{t \rightarrow +\infty } -\log(1-W (t)) =\infty$ with  probability one \citep{walker1997}.
 Dirichlet processes  constitute   a subset  of BS models.
But unlike   the Dirichlet process,  the BS prior is
conjugate with respect to  right  censored data \citep{walker1997}.  
If  $Z=\{ Z_i \}_{i=1}^n$  
is an independent,  right-censored sample from  a distribution $W$ and $ W \sim BS(W | V_0, c)$, 
%$F \sim BS(F | c, H_0),$
then   $p(W | Z) = BS(W | V_n, c_n)$
is again a BS  with closed from 
expressions  for the posterior mean $V_n$ and  uncertainty parameter $c_n$   as described in \citep{walker1997}.  
An advantage of using the BS prior  is that, conditionally on  right censored data,
the posterior distributions are available in closed form, and 
the summaries $\beta$ and $\theta$  %  \footnote{replace  with  "$\beta$ and $\theta$"  to  avoid notation conflicts }
% \footnote{SV: yes,   notation introduce before}
can be easily   simulated from the posterior. %\footnote{use $D$  instead of $F$}
 %  from the Beta-Stacy posterior.
%Posterior computations are carried out using the Monte-Carlo algorithm described in the supplementary material.
%
%equal to
%%
%\begin{align}
%1-
% \prod_{\ell:  t_{ \ell}  \leq t } \Bigg[  1 - \dfrac{ H_0 \{  t_{u,\ell}  \}  + N (t_{u,\ell}) } { R_m(t) +  c(t) G( [t, +\infty)  } 
%\Bigg ]
% \times \exp \bigg \{ - \int _{0} ^ t    \dfrac{ d H_{0,c} (s) } {  R_m(s)  + c(s)}  \bigg   \},
%% %
%\end{align}
%%where 
%$w_{n}(t) = \dfrac{ 1 }{1 +  R_n(t) / \{  c(t) G( [t, +\infty) \}  }$, 
%and 
%$c_m (t) $ equals $\dfrac{    c(t)  H_0([t,+\infty) ) +   R_m(t+)  }  { H_m([t,+\infty) }$ for every $t\geq 0$,
%where  $R(t)_m =\sum _{1 \le \ell  \leq m}  I( Z_\ell \geq t)$ and $N_m(t) = \sum _{1 \le \ell  \leq m}  I( \Delta_\ell = 1,   Z_\ell=t).$
%
% efficacy and

 When  the   same therapy  is  evaluated  at  
    decreasing dose levels $k=1, \cdots, K$,  
    one  can assume that  toxicity is non-increasing   with $k$ and 
 %In this case  
enforce monotonicity   of  the parameters $(\beta_k)_{k=1}^K$     across treatments $k$ with probability one
 % i.e  $p(\theta_{k}\ \geq \theta_{k+1})=1$
 % and $p(\beta_{k}\ \geq \beta_{k+1})=1$,
by  multiplying   independent  prior  distributions for $(G_k)_{k=1}^K$
and  the  indicator  function of the event  $\{  \beta_1 \leq  \beta_2 \ldots  \leq  \beta_K\}$, 
\begin{align}\label{Post:Mon}
p(G_1, \cdots, G_K) \propto I( \beta_1 \leq  \beta_2 \ldots  \leq  \beta_K) \prod_{k=1}^K BS(G_k | G_0, c).
\end{align}

%
%$\footnote{$BS(F_k ; W,c)$  might be better  than $BS(F_k | W,c)$}
%

In our oropharynx cancer study the treatment arms $k$ are  decreasing radiation doses.
We  can therefore also assume that  the efficacy  is monotone non-increasing with respect to the $k=1, \cdots, K$  dose levels.  
For $(F_k)_{k=1}^K$, we therefore multiply, similar to (\ref{Post:Mon}),  the   independent  BS prior probabilities   by the indicator   
function  $I(\theta_1\geq \theta_2 \geq \ldots  \geq \theta_K)$.
For clinical trials  that  evaluate different therapies, 
the  assumption  $\theta_k \geq \theta_{k+1}$
may not be appropriate, and can be easily removed from the model.  
\section{HPV-associated oropharynx cancer}\label{Sec:Simulation}
In Section \ref{Sec:Simulation:Study:Effiacy} 
we discuss the sensitivity of  the trial operating characteristics  to  the design  parameters. 
We then evaluate in Sections 
\ref{Sec:Simulations:Effiacy:Tox} 
and 
\ref{Sec:Simulation:Study:Effiacy:Parametric}
the 
de-intensification designs with efficacy  outcomes %(Sections \ref{Sec:Design:EfficacyEndpoint}),
and with efficacy and toxicity co-primary outcomes. %(Sections \ref{Sec:Design:Two:Endpoints}). 
%in HPV-associated oropharynx cancer.

\subsection{De-escalation design with efficacy endpoints}\label{Sec:Simulation:Study:Effiacy}
We evaluate  the operating characteristics of the design in Section \ref{Sec:Design:EfficacyEndpoint}  for different    
values of $\mathcal S=(s_I, S_I, S_{NI}, m_{NI}, m_{I})$. %
We  focus on a study that evaluates a single de-intensified treatment  
with maximum sample size of $n_{\max}=100$ patients,
an average enrollment of  $5$ patients per month, 
$t_{FU}=12$ months follow-up, 
and  $\theta_k$ is the 24-months RMST $\theta_k = E[ \min(Y_i, 24)| C_i=k]$. 
The 24-months RMST of the  SOC    is $\theta_0 =22,$ (similar to cisplatin+IMRT in HPV-associated  oropharynx cancer)   the null hypothesis is $\theta_1 \leq 20$, $\Delta=2$. 
%and  $\Delta_1=\Delta=2$.
%
The prior is centered at an exponential distribution with RMST $20$ and $c_0(t)=10.$ 
We consider five scenarios with  RMST of arm 1 equal to $19,20, 21, 21.5$ or $22$ months 
and a   targeted type I error rate of  $\alpha=0.1$  for  our   phase II de-escalation study.

We calibrate the parameter  $s_I$ of the futility boundary $b_I(\cdot)$ 
so that, assuming  an exponential  outcome distribution,
approximately a proportion $p_I=0, 0.1, \cdots, 0.9$ 
of the simulated trials are stopped early for futility when
$\theta_1=\theta_0-\Delta$.   
Algorithm S1 in the supplementary material describes this calibration.
 
\begin{figure}[htbp] 
\center 
\includegraphics[scale=.99] {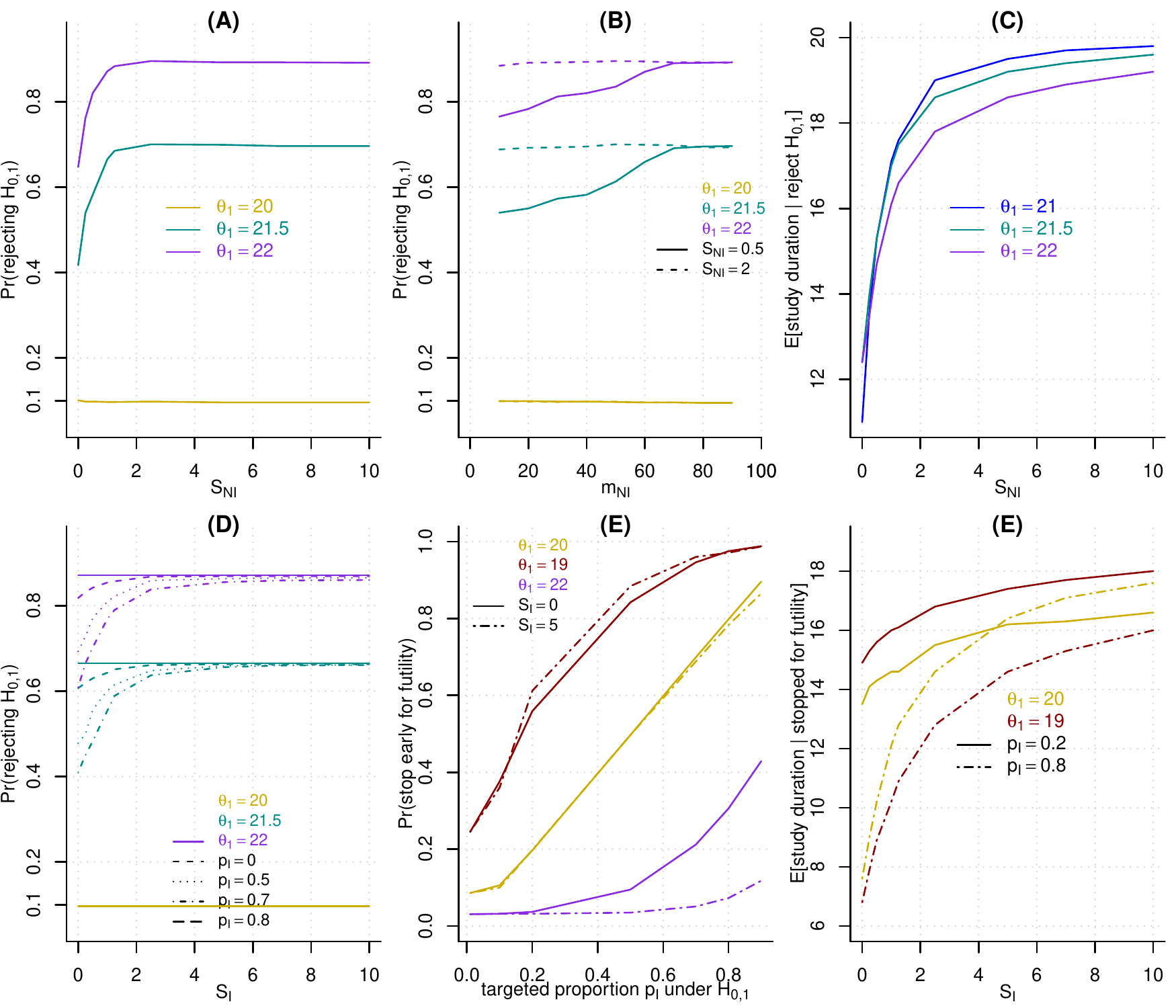}
\caption{
Operating characteristics of the  de-intensification design 
in a single-arm study, with maximum sample size  $n_{\max}=100$ and $m_{F}=0$ 
for different design parameters  $(p_F, S_F, S_{NI}, m_{NI})$.  
In all Panels, except panel two in  the top row, $m_{NI}=50$.
The first row shows,  for $p_F=0$ and different  values  of $S_{NI}$ and  $m_{NI}$, the power and  the average study duration.
The second row shows, for $S_{NI}=1$ and  different  values of  $S_F$ and $p_{F},$ the power, the probability of stopping the study early for futility and the average study duration. 
}\label{Fig:Sensitivity1}
\end{figure}

Figure \ref{Fig:Sensitivity1}  summarizes selected operating characteristics of the Bayesian de-intensification design of Section \ref{Sec:Design:EfficacyEndpoint} across different values of 
$S_j, j=I, NI$, 
$p_I$ and $m_{NI}$.
Panels A and C   of Figure \ref{Fig:Sensitivity1} indicate 
that, as  expected, increasing values of $S_{NI}$ lead to larger power, 
but lead also to an increasing average time before effective de-intensified therapies are declared non-inferior.
Similarly, for boundaries   $b_{NI}(\cdot)$  with shape  parameter $S_{NI}$ close to $0$, 
increasing values of $m_{NI}$ lead to an increase in power (Panel B). 
Symmetrically,  increasing values of $S_I$ lead to higher power when large target proportions $p_I > 1/2$ are used (Panel D),  
but also  to  an  extended average time required to stop inferior arms for futility  (Panel E).
Panels E and F show that when large   $p_I > 1/2$ are used, 
the probability of stopping a non-inferior treatment for  futility  
and  the power reduction can be  large  unless values of $S_I \geq 3$ are used.

\subsection{HPV-associated oropharynx cancer}\label{Sec:Simulation:Study:Effiacy:Parametric}

\begin{figure}
\includegraphics[scale=1] {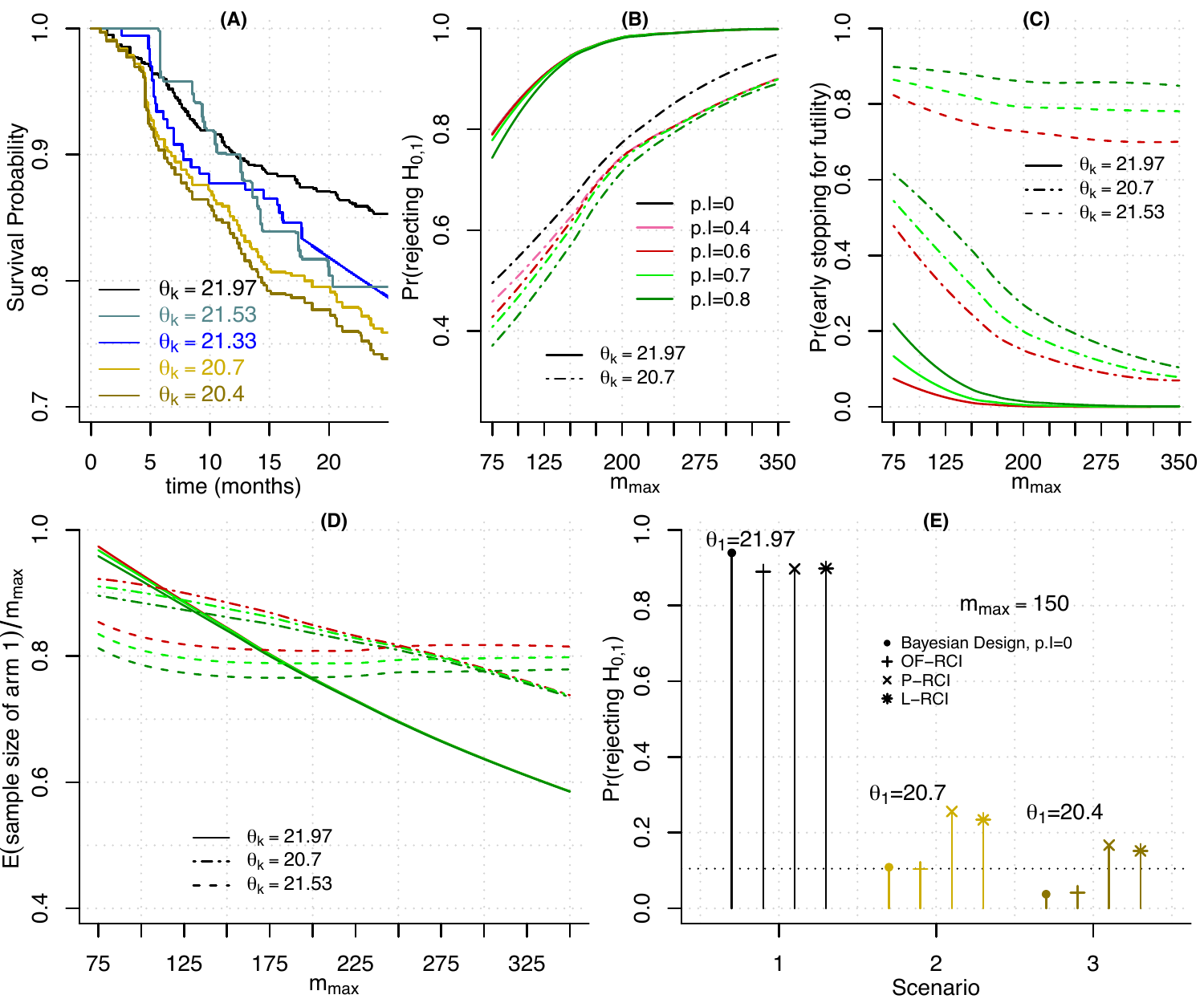}
\caption{Panel A shows PFS Kaplan-Meier curves extracted from four recent de-intensification studies (RTOG 1016, DeEscalate, Optima and E1308) which we use to generate outcomes $Y_i$.
Panels B to D show selected operating characteristics of treatment $k=1$ for a maximum sample sizes $m_{\max} =75, \cdots, 350$. 
Panel E shows, for $m_{\max}=150, p_I=0, S_{NI}=6, m_{NI}=50$, power  when  $\theta_1=21.97$ (black bars),   $\theta_k=20.7$ (yellow bars)   
and $\theta_k=21.97$ (brown bars) for the proposed Bayesian design  and the three RCI methods 
with O'Brien-Fleming\citep{o1979multiple},  Pocock  \citep{pocock1977group} 
and linear spending functions \citep{reboussin2000computations} (OF-RCI, P-RCI and L-RCI).	}\label{Fig2:ResimulateBySampleSize}
\end{figure}

We apply the design of Section \ref{Sec:Design:EfficacyEndpoint}  
retrospectively to four recent clinical studies in HPV-associated oropharynx cancer.
We extracted published PFS distributions  $F_k$ (Panel A of Figure \ref{Fig2:ResimulateBySampleSize}) from the  recent de-intensification studies
RTOG 1016, DeEscalate, Optima and E1308 \citep{gillison2019, mehanna2019, seiwert2019optima, marur2017e1308}, 
using the software DigitizeIt \citep{DigitizeIt}.
RTOG 1016 is a large randomized
 (849 patient)  phase III  study, whereas  the remaining three studies 
where  smaller single arm studies. 
The IMRT+{\it Cisplatin} SOC (black curve in Panel A)  has an estimated  24-months RMST of $\theta_0 = 21.97$ months.

We consider a study with  two  de-intensified therapies, an average 
of 5 enrollments per month, 
monthly interim analyses,  $t_{FU} = 12$  months follow-up time  and 
 null hypotheses 
   $\theta_k \leq 20.7$ ($\Delta = 1.27$)
   are tested at an  $\alpha=0.1$ level. 
For the Bayesian design we used  $(S_I, S_{NI},  m_{NI}, m_I)=(5,6,50, 0)$. 
In each of the scenarios that we consider below,  
we use a different pair of distributions from  Panel A  of Figure \ref{Fig2:ResimulateBySampleSize}, to sample 
PFS  outcomes for the 1st and 2nd  treatment.
Blue and black survival functions have RMSTs $\theta_k$ that are  non-inferior to the SOC,  
%(although most clinical investigators would not consider reductions on 24-months RMSTs of 0.5-0.8  months desirable), 
whereas the remaining two distribution functions (yellow and brown) have inferior RMSTs.  

%\footnote{in panel A  the  reported $\theta_k$  values  are  wrong}

We initially determined a sample size $m_{\max}$, assuming $F_1=F_0$, 
to achieve approximately a  power of 90\% for the first $k=1$ experimental arm,
enforcing different early  stopping probabilities  $p_I$ 
under the null hypothesis  when 
$\theta_1=20.7$ %\footnote{conflict you previously wrote "  $\mathcal H _{0,k}$ of $\theta_k \leq 20$ "}
(Panel B of Figure \ref{Fig2:ResimulateBySampleSize}).
For  $\theta_1=\theta_0$, the power shows   little  sensitivity to the choice  of  $p_I$. 
Whereas  with $\theta_1=21.5$ the power  
varies  substantially  with $p_F > 0.6$. 
Based  on the Monte-Carlo calculations in Panel B of Figure \ref{Fig2:ResimulateBySampleSize}, we select  $m_{\max} = 150$ and use $p_I=0.7$.

\medskip
{\it Comparator designs. } 
We compare the Bayesian design to alternative  
 de-intensification designs 
with  different combinations of  testing and futility stopping rules 
\citep{jennison1989interim, o1979multiple, pocock1977group, reboussin2000computations}. 
At each IA, similar to the Bayesian design, 
these  designs may  declare a treatment  $k$   
non-inferior  and start  evaluating  arm $k+1$, 
or declare treatment $k$ inferiority and stop the study. 
 Futility and non-inferiority IA are conducted monthly,  starting after $m_{\min}=50$ enrollments.
  
Non-inferiority is tested  in the comparator designs %analyses  (rejecting $\mathcal H_{0,k}$) 
using  the {\it Repeated Confidence Interval (RCI)} method 
\citep{jennison1989interim, gillison2019}.  
At each interim and final analysis $t$, 
we   estimate the RMST
$\widehat \theta _k$ from the Kaplan-Meier estimate   $\widehat F_k,$ 
 $\widehat{Var}(\widehat \theta _k)$ via   bootstrap,  
 and a   $(1-\alpha_t)$-confidence interval    $[ L_{\alpha_t}, +\infty)$   
for $\theta_k$  obtained  using the  asymptotic normal distribution  of   $\widehat \theta _k$ \citep{zhao2016restricted}.
$\mathcal H_{0,k}$  is then rejected if   $  L_{\alpha_t}>20.7.$   
The values $(\alpha_t)$ are  error-spending functions \citep{reboussin2000computations}
targeting  an overall %multiplicity-adjusted   
$\sum_t \alpha_t= 0.1$ type I error rate.  
We consider O'Brien-Fleming\citep{o1979multiple},  Pocock  \citep{pocock1977group} 
and linear functions  \citep{reboussin2000computations} (OF-RCI, P-RCI and L-RCI).

For  futility IAs  in the  comparator study designs 
we consider  three frequently used stopping rules.
(1) At each IA $t$ we compute a p-value for the `'null'' hypothesis ($\theta_k \geq 21.97$) using a normal approximation for the distribution of 
$\widehat \theta _k$, 
and stop the study  if the p-value $\leq 0.0025$ as suggested in \citep{freidlin2010general, gillison2019}. 
(2) Alternatively, \citep{lachin2009futility} suggested to use a  p-value $\leq 0.05$ cut-off. 
%   
%we consider a  $95\%$ confidence interval $(-\infty,  U_{95\%,t}]$ for  $\theta _k$ 
%and stop the study when $ \theta_0=21.97 >  U_{95\%,t}$ \footnote{I REPEAT AGAIN  2 and  1 are  equivalent procedures  with different  
%parameters }
(3) The last  futility rule uses the RCI 
 \cite{freidlin2010general} and stops the study  at  IA  $t$
if the   $(1-\alpha_t)$ 
confidence interval $(-\infty, \widehat U_{t}],$ with overall $\sum_t \alpha_t=0.025$ as 
suggested in \citep{freidlin2010general},
doesn't  include $\theta_0=21.97$.

%Before we investigate the designs operating characteristics across all eight scenarios below, 
We first  compared type I error rates and power of the  designs 
for the first de-intensified therapy ($k=1$) in three scenarios with $\theta_1 = 20, 20.7, 21.97$ and $m_{\max} = 150$ patients (Panel E  of Figure \ref{Fig2:ResimulateBySampleSize}).
To simplify the evaluation, we don't consider interim futility analyses in these three scenarios. 
RCIs  with Pocock  \citep{pocock1977group} and linear spending functions \citep{reboussin2000computations} 
do not control  type I error rates at the targeted  $\alpha=0.1$ level 
with  empirical   type   I error rates of   $0.26$ and $0.23$ across 10,000 simulations.
O'Brien-Fleming boundaries have type I error rates nearly identical to the nominal $\alpha$ level. 
Panel A of Figure S1 shows that the normal approximation of the  RMST estimates $\widehat \theta_k$ in the {\it RCI}  
is not  accurate  for  the initial  IAs, which  lead to  these inflated error rates,
% and underestimates the right tail of the distribution of  $\widehat \theta_k$, \footnote{very  unclear "and underestimates the right tail of the distribution of  $\widehat \theta_k$" cut}
whereas the approximation becomes  better towards the end of the study (Panel B). 
O'Brien-Fleming boundaries are significantly more conservative  during the initial  IAs 
than the linear and   Pocock's boundaries,  and hence O'Brien-Fleming boundaries are  
less affected by these  approximation errors.  Therefore,
for the remaining comparisons   we use     O'Brien-Fleming RCI-boundaries 
and  the  three futility stopping rules described above (RCI-F1, RCI-F2 and RCI-F3).

\begin{figure}[htbp]
\includegraphics[width=17cm, height=17cm] {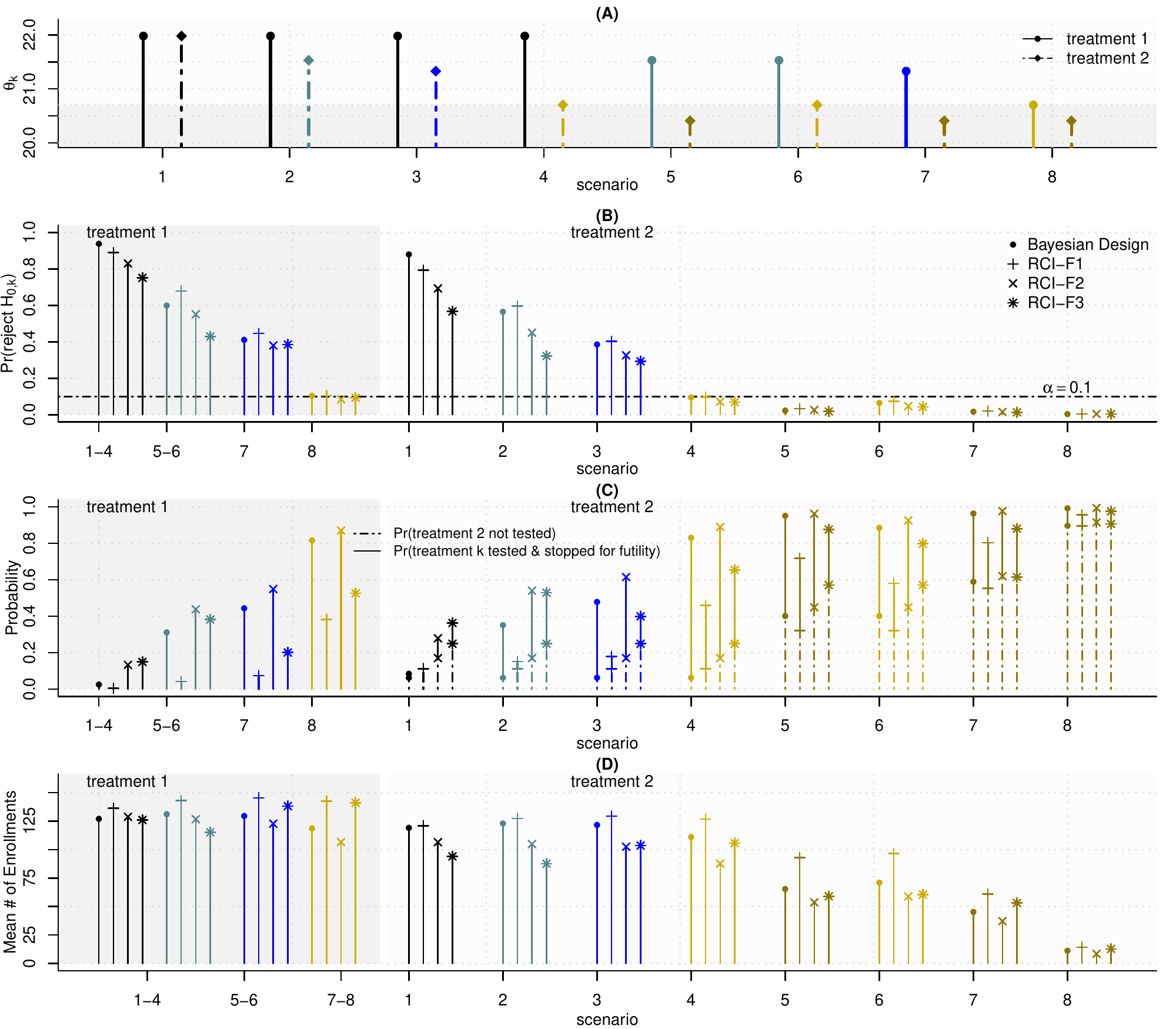}
\caption{
Operating characteristics of the de-intensification designs for a two-arm study with efficacy primary endpoint 
and a maximum of $m_{\max}=150$ patients   for each treatment.
Panel A summarizes the 24-month RMSTs $\theta_1$ and $\theta_2$ (y-axis) in each of the eight  scenarios (x-axis) that we consider.
Panel B shows the power for the Bayesian design  and the three alternative designs (RCI-F1, RCI-F3 and RCI-F3).
Panel C shows probability of stopping  treatment 1 and 2  at an IA for futility (solid vertical line), 
%given that the study started testing treatment $k=1, 2$ 
and  probability that  the 2nd treatment is not tested due to early termination of the study (dashed vertical line).
%For the 2nd treatment, the probabilities of early stopping for futility are computed conditionally on the event that the study starts  testing  arm 2.  
Panel D shows the average enrollment  of the  two-arms  trial  on   the 1st and 2nd de-intensified treatments.
}\label{Fig3:TwoArmScenarios}
\end{figure}

\medskip
{\it Operating characteristics for  arms $k=1,2$ .} Panel A of Figure \ref{Fig3:TwoArmScenarios} summarizes the 8 scenarios that we consider in the two-arm  study. 
For each scenario (x-axis) the first  and second vertical  bar  indicate the RMSTs $\theta_1$ and $\theta_2$ that we consider with  distributions  $F_1$ and $F_2$ selected from   Figure    \ref{Fig2:ResimulateBySampleSize}  (same colors).
%the corresponding  $F_k$ are indicated by the same color as the horizontal lines).
%The first three  scenarios correspond to cases   with non-inferior 
Treatments $k=1, 2$ are  non-inferior in the first three  scenarios,
whereas the second de-intensified treatment  is inferior  in  the last four scenarios. 
  
In scenario 1, both de-intensified treatments are non-inferior to the SOC with identical RMSTs $\theta_0=\theta_1=\theta_2$.
The Bayesian design has 93.8\% and 87.9\% power to declare the two arms non-inferior, 
compared to 88.9\% and 80\% for the RCI-F1 design, see Panel B of Figure \ref{Fig3:TwoArmScenarios}.  
The remaining two designs RCI-F2  and RCI-F3 have  lower power 
(73.8\%  and 54.1\% for RCI-F2 and 75.1\% and 56.6\% for RCI-F3), respectively. 
The power  in Panel B of Figure \ref{Fig3:TwoArmScenarios}  
for the second experimental arm, is defined as the probability that the study starts testing treatment  2 and rejects $\mathcal H_{0,2}$ at final or IAs. 
Panel C shows, for both,  arm $k=1, 2$ the probability that the study started testing treatment $k$ and stopped  treatment $k$ early for futility at IAs 
(solid vertical bar). 
For the 2nd treatment we also show the probability that the study does not start testing the treatment (dashed vertical bar). 
For instance, for the Bayesian design in  scenario 5, 
the   inferior 2nd treatment is not tested or it is stopped early for futility with probability 0.95.  
Here the study does not start testing the treatment with probabilities 0.40 and is stopped early for futility with probability 0.45.
%
%Panel C indicates that the reduction in power of RCI-F2  and RCI-F3 compared to RCI-F1 is largely due to different  frequencies  of  early stopping.
%\footnote{cut " Panel C indicates that the reduction in power of RCI-F2  and RCI-F3 compared to RCI-F1 is largely due to different  frequencies  of  early stopping.
%"  obvious  that  is  the only  difference  between designs}
%
Panel C  shows that the futility stopping rule of RCI-F1 
leads to a low probability of stopping inferior treatments (scenario 8, $\theta_1=20.7$) early for futility (38.3\%) compared to the RCI-F2 (93.2\%) and RCI-F3 (52.6\%) 
and the proposed Bayesian design (81\%).
This  leads in scenarios 5 and 7, where the first de-intensified treatment is non-inferior, but $\theta_1$ is  close to  $\theta_0-\Delta=20.7$ ($\theta_1=21.53$ and $21.33$),
 to a slightly larger power of the RCI-F1 compared to the remaining designs.

Scenarios 4 to 8, where the second de-intensified treatment is inferior to the SOC, 
shows the benefit of testing experimental arms sequentially. 
For instance, if the first experimental arm is inferior ($\theta_1=20.7$, scenario 8), 
all designs  start testing the inferior 2nd   treatment  with less than $10\%$ probability ($10\%$ for RCI-F1, RCI-F3 and the Bayesian design, and 7\% for RCI-F2).

\subsection{The use of efficacy and toxicity co-primary outcomes}\label{Sec:Simulations:Effiacy:Tox}
We  consider testing   non-inferior survival and   toxicity reductions
during the de-intensification study. % using the design of Section \ref{Sec:Design:Two:Endpoints}.
We assume the same enrollment rate  (5 enrollments per months),  
prior model for  $(F_k)_k$, $t_{FU}=12$ months and 
a 24-months RMST of $\theta_0=29.97$ for the historical SOC as in the previous section,
and a  24-months RMST for the 1st AE of grade $\geq 3$ of $\beta_0=12.49$ months (estimated from published data of RTOG-1016).
The null hypothesis  that we test is 
$\mathcal H_{0,k} =\{ (\theta_k, \beta_k)  \in \mathbb R ^ 2:  \theta_k \leq 20.7  \text{ or }  \beta_k \leq 12.49 \}$.
We consider again simulation  scenarios with  distributions $F_k$ (and $\theta_k$)
  identical to  Kaplan-Meier curves  in Panel A of Figure \ref{Fig2:ResimulateBySampleSize}.
Simulation scenarios include  exponential  distributions $G_k$ for the time $X_i$ until the 1st  grade-3 AE
with 24-months RMST equal to 
$\beta_k=12.49$
 or $14.5$ months.%, for toxicity endpoints $X_i$.

We use  $\Delta = 2,$ %\footnote{SV: No, $\Delta$. There is no extra parameter $\Delta_H$ }
$\Delta_L = 1$ in (\ref{ET:Stop2:H0}) and $\Delta_\beta=0$. 
The shape parameters $S_j=5$ for the futility and toxicity boundaries  $b_{I}, b_T$
%\footnote{$b_{F}$ or $b_I$??} 
and $B_T$ 
%\footnote{can you plot boundaries in supplementary material?  this  would make  transparent  that many parameters  translates only 
%in a a few  functions activated at  specific  time  intervals (x-axes  "enrolled patients" y-ax "posterior probability thresholds" ) }
 in (\ref{Design1:boundary}), 
$S_{NI}=6$ for $b_{NI}$, and we required $60$ 
%\footnote{SV: $m_T$  introduced in table but not in text}
assignments before applying (toxicity, futility and non-inferiority) stopping rules.  
We tuned the scale parameters $s_j$ in (\ref{Design1:boundary})  for  $b_I, b_T$ so that with probability $0.5$ 
%\footnote{SV: $p_T$  introduced in table but not in text}
inferior treatments or treatments that do not reduce toxicities are stopped early at IAs.

\begin{figure}[tph]
\includegraphics[scale=.9] {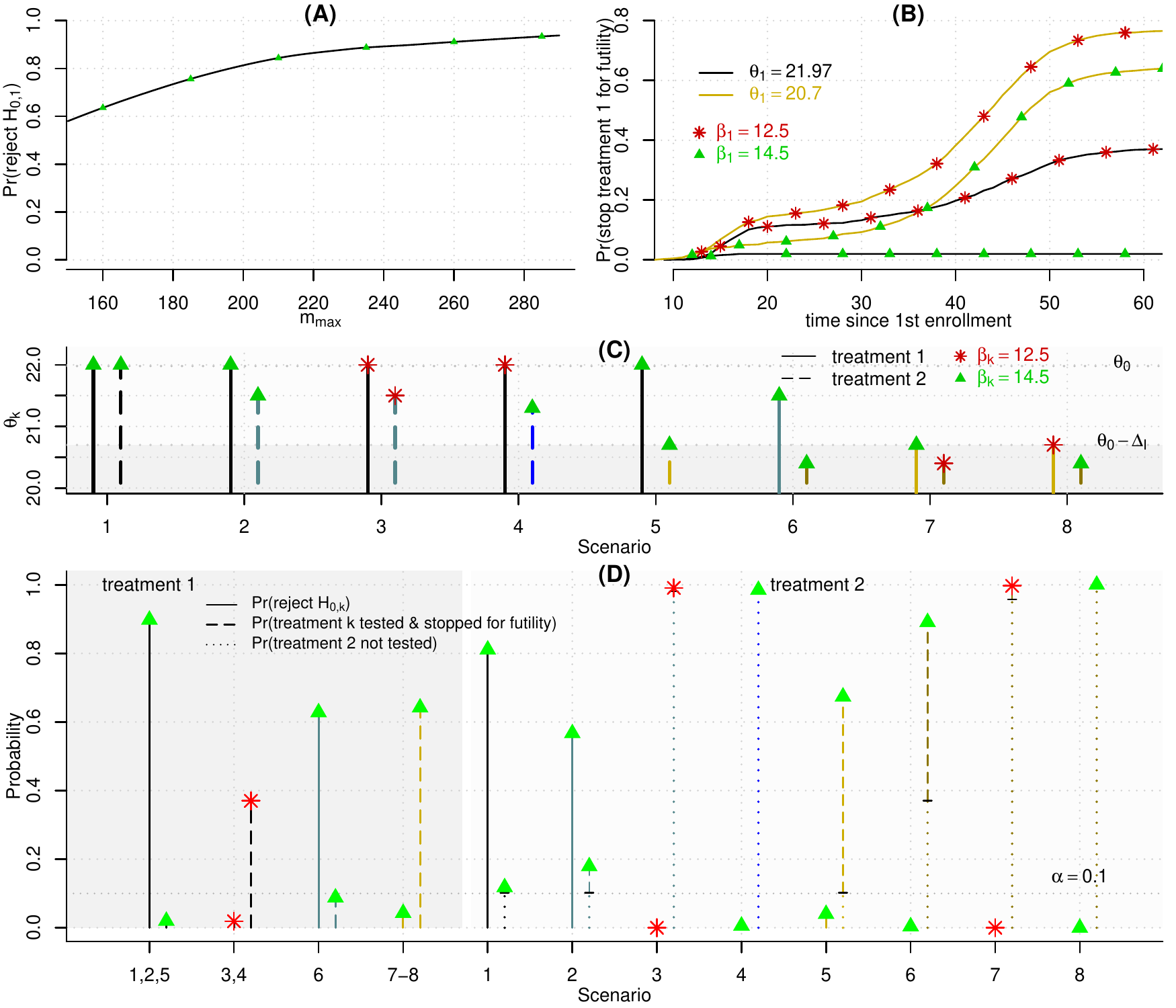}
\caption{Operating characteristics of Bayesian de-intensification designs for a study with efficacy and toxicity co-primary endpoints.
Panel A shows the power for  treatment $k=1$ when $\theta_1=\theta_0$ and $\beta_1=\beta_0+2$
with maximum sample size  $m_{\max}$ between $150$ and $300$ patients.   
Panel B shows the probability of stopping treatment $k=1$ for futility (either for inferior survival or low evidence  of reduced toxicities) 
when $\theta_1=\theta_0$ (black curves) and $\beta_1=\beta_0,\beta_0+2$ (red cross and green triangles)
and 
when $\theta_1=\theta_0-\Delta=20.7$ (yellow curves) and $\beta_1=\beta_0,\beta_0+2$ (red cross and green triangles)
for a study with maximum sample size  $m_{\max}=250$ patients. 
Panel C summarizes the 24-month RMSTs $(\theta_k, \beta_k), k=1,2$ in each of the eight  scenarios (x-axis) that we consider.
The  vertical bars (y-axis) indicate $\theta_k$, whereas  green arrows ($\beta_k=14.5$) and  red stars ($\beta_k=12.5$) on top of the vertical bars indicate toxicity parameters.
Panel D shows, for both treatments, the power (solid bar), the probability of stopping  treatment evaluation early for futility at IAs (dashed bars), and     the
probability that  the 2nd treatment is not tested due to early termination of the study (dotted  bars).
}\label{Fig:ToxEfficacy}
\end{figure}

We first determined for the 1st arm the power  of the Bayesian design 
for a  maximum arm-specific sample sizes $m_{\max}$ within the range $150$ to $300$  patients
(Panel A of Figure 
\ref{Fig:ToxEfficacy})  when  $\theta_1=\theta_0$ and $\beta_1=\beta_0+2$.
With a targeted   $\alpha=0.1$ level, 
the design requires approximately 197  and 250 patients to achieve  80\%  and 90\% power, respectively.

We then considered a two-arm de-intensification study  ($K=2$) with maximum overall sample size per arm of $m_{\max}=250$
patients 
and evaluate the operating characteristics of the proposed Bayesian  design in 8 scenarios  that 
are summarized in Panel C 
of   Figure \ref{Fig:ToxEfficacy}.
PFS parameters $\theta_k$ are represented by vertical bars in Panel C 
(solid and  dashed bars for $\theta_1$ and $\theta_2$, 
the colors are consistent with 
the distributions $F_k$ in Panel A of Figure \ref{Fig2:ResimulateBySampleSize}). 
Toxicity parameters are indicated by the green triangles ($\beta_k=14.5$) and red stars ($\beta_k=12.5$) on top of the vertical bars.

Panel B of   Figure \ref{Fig:ToxEfficacy}  shows the benefit of  interim monitoring of efficacy and toxicity endpoints.
The figure shows, for the first treatment, the cumulative probability of stopping the treatment for futility  (y-axis), i.e. for inferiority or toxicity, by time $t$ since the first enrollment (x-axis) for four  scenarios (scenarios 1, 3, 7 and 8: all combinations of $\theta_1 = \theta_0 - \Delta, \theta_0$ and $\beta_1=\beta_0, \beta_0+2$).   
For instance,  if the treatment is inferior with  $\theta_1 = \theta_0 - \Delta=20.7$ but  reduces toxicities ($\beta_1= \beta_0+2$), then
64\% of all simulated de-intensification trials are stopped early for futility at IAs (scenario 8, golden curve with green triangle).
In comparison, if the treatment  fails to reduce toxicities  ($\theta_1=20.7$ and $\beta_1= \beta_0$), then
the treatment   is stopped early for futility  in 77\% of the simulations (scenarios 7, golden curve with red stars).

Panel D
 of   Figure \ref{Fig:ToxEfficacy}  shows  for both treatments  the probability of
rejecting $\mathcal H_{0,k}$ (power, solid vertical bars), and  
the probability that the study evaluates  treatment $k$ and stops this arm  early for futility (inferior survival or insufficient reduction of toxicities) at IAs (dashed vertical bars).
As before the power for the 2nd treatment is defined as the probability that the study starts testing treatment  2 and rejects $\mathcal H_{0,2}$.
For  the 2nd treatment Panel D shows also the probability that the 2nd treatment is not tested due to early termination of the study (dotted vertical bars).
%and the probability of stopping treatment 2 for futility at IAs is computed conditional on the event that the study starts testing  treatment 2.  
%
If both treatments ($k=1,2$) extend % \footnote{extend?}
the the RMST of the 1st AE  by two months ($\beta_1= \beta_2 = 14.49$) 
compared to the SOC and 
have identical survival outcomes ($\theta_1= \theta_2 =  21.97$) 
the 1st and 2nd de-intensified treatment have  $90\%$ and $82\%$ power (scenario 1).
In comparison, with a moderate non-inferior treatment effect of $\theta_k=21.53$ in scenario 6 the power for  the 1st treatment decreases to 
$63\%$.

Similar to the Bayesian design with efficacy primary endpoint  of Section \ref{Sec:Design:EfficacyEndpoint}, %\footnote{add  (subsection xxx)}, 
scenarios 7 and 8 indicate the advantage of 
testing the 1st and 2nd treatment sequentially one after the other. 
If the first treatment is inferior, 
the second  treatment $k=2$ is  tested in 4\% (scenario 7, $\beta_1=14.49$) or in less than $1\%$ (scenario 8, $\beta_1=12.49$) of all simulations.
Scenario 4 shows one limitation  of the  design, here both treatments have non-inferior survival 
but only the 2nd treatment reduces toxicities, $\beta_1 = 12.49$ and $\beta_2 = 14.49$. 
In this case the study is terminated early without testing the second treatment in $98\%$ of all simulations.

%\footnote{what  you  call  arrow are triangles}

\section{Discussion}\label{Sec:Disuccion}
%- recent interest in DI trials, especially in head and neck cancer and breast cancer
%
%- trials differ from conventional trials that test efficacy 
%-- larger sample sizes are required to test treatments at standard ni-margins
%-- investigators tend to use large margins to reduce the sample size, 
%-- what counts are survival effects nearly identical to the SOC while reducing toxicities 
%-- Whereas in efficacy trials most ineffective arm have survival similar to the SOC
%-- ineffective de-intendified in trials with large NI-margins reduce patients survival substantially 
%-- for instance RTOG 1016 and DeEscalate 
%
%- X et all - list X ongoing trials, 
%-- many of them do not consider explicit futility stopping rules and test only one of efficacy or toxicity 
%- for instance RTOG 1016 and DeEscalate test only efficacy or totociity 
%
%- present trial designs testing effiacy and toxicity
%-- explicitly select parameters to control trade-offs between power and exposing patients to inferior or toxic treatments
%-- flexible and robust Bayesian
%-- if two or more related de-intensified are tested we propose to investigate the most promising treatment first 
%-- decision to evaluate the second treatment can be made after the first treatment was succesfully or unsucessfully evaluates
%-- similarly once can decide  to use less stringent futilitu stopping rules for efficacy 
%
%- frequentist designs, typically based on normal approximations, are difficult to apply in settings where 

%In oncology % - especial in malignancies with large median survival times such as head-and-neck and breast  cancer  - 
There has been a recent interest in %the development of Recent studies  discussed  the  importance 
  %of   
  developing de-intensified anti-cancer treatments   with similar survival  as the current  SOC and  reduced AEs. 
%and to increase patients quality of life. % during and after cancer treatment. 
\citep{elrefaey2014, mirghani2018treatment} identified 12 de-intensification studies in  oropharyngeal cancer that are currently ongoing or  that  recently reported results.

Compared to traditional superiority trials, which test  superiority of  experimental treatments compared to the SOC,
%that test improvements in survival of experimental therapies relative to the current SOC:
demonstrating (i)  similarity  in survival between de-intensified  treatments and the SOC, and (ii) reductions in  AEs %to justify potential reductions in average survival. 
require 
large sample sizes.  % than efficacy studies, 
Investigators often select large NI margins  to reduce the  size  of the  study  \citep{ventz2019lessons}. 
%reduce sample sizes required for a target power. 
%If the  margin $\Delta$ is large,  inferior de-intensified treatments  reduce  patients average survival significantly relative to the SOC.
%which is very different from efficacy-studies where ineffective treatments have typically survival outcomes similar to the SOC. 
%
Recent   results  in oropharyngeal cancer  
showed  that many de-intensification treatments fail  to reduce toxicities and have  inferior survival compared to the SOC \citep{ gillison2019, mehanna2019}.  
As discusses in \cite{ventz2019lessons},
many of these studies 
(i) evaluate  only  survival or  toxicity, 
(ii) do not have explicit futility early stopping rules for survival   and  toxicity  endpoints, and
(iii) tend to use  conservative  early stopping rules to avoid  power reductions.

Motivated by a  study at our institution, which tests  two  dose-reduced treatments, %(i.e.  {\it  intermediate}  and  {\it low} dose levels),
we proposed a Bayesian design for multi-arm de-intensification studies.
The design tests non-inferior survival and toxicity reductions sequentially using a Bayesian non-parametric model.
We proposed futility stopping rules  to  monitor both endpoints. 
The design parameters %$(p_{F}, p_{T})$ 
can be tuned to  calibrated trade-offs  between power and the  probabilities of stopping   arms early for inferior  survival  or insufficient evidence of toxicity reductions. 
In oropharynx cancer, where survival rates five years after      {\it IMRT+cisplatin}  treatment are $>90\%$, 
%of the patients treated with standard     {\it IMRT+cisplatin} survive more than 5 years and 
the number of observed OS and PFS events during the trial are typically small. 
Standard frequentist methods based on large-sample normal approximations can perform  poorly in this setting and  
the Bayesian approach is an attractive alternative. 
%($p_{T}$).
%The Bayesian approach is attractive in the de-intensification setting,
%because   we can leverage  available  data  on the  SOC.
%Additionally, in settings with good with low  
% because  event rates during the trial tend to be low,  predictions  become essential,
%and  standard frequentist methods based on large-sample normal approximations  perform  poorly and   can  (as  shown in our simulation study) inflate  the type I error rate.

De-Intensified treatments in our design are tested one at the time, 
starting with the treatment with the highest dose-level. 
This controls  the  number  of patients exposed  to  inferior treatments.  
%because whenever arm $k$ is inferior to the SOC and is stopped at IA for futility, 
%no patient will be exposed to treatment $k+1$ which is a further dose reduction compared to arm $k$.
%
As indicated in Section \ref{Sec:Simulations:Effiacy:Tox},	one limitation of the Bayesian design is that it could
   stop the first non-inferior arm  ($\theta_1>\theta_0+\Delta$) because of  frequent  toxicities  and 
terminate  the study, but  the 2nd treatment may potentially   reduce toxicities  ($\beta_2>\beta_0$)  with  non-inferior
survival  ($\theta_2>\theta_0+\Delta$). 
The design can  be  modified,  to account for  this  limitation,  with a  definition of  the  decision to  terminate the study  or  not  
that  distinguishes  between negative  results  for  arm  1   due  to  toxicities   or due to  PFS  data.

 We focused  on non-controlled phase II de-intensification studies, but
the design could be  modified to include a concurrent control arm. 
If the SOC  is associated with  severe  toxicities,  
it may  be unethically to continue  the  assignment of  patients to the SOC  after   the  null  hypothesis   $\mathcal H_{0,1}$ has been rejected.
In this case the 1st de-intensified treatment could  be utilizes as the  {\it new control arm}  for testing the 2nd de-intensified experimental treatment.
%\footnote{last  2  sentences:  not  well written  and  not  sufficient   to  cover  the  promise  of  discussing  the control arm  option  early in the  paper}

{\small \bibliography{NI_ref.bib} }

\end{document}